\documentclass{PoS}

\title{Possible non-prompt photons in $pp$ collisions and their effects in $AA$ analyses}

\ShortTitle{Possible non-prompt photons in pp collisions and their effects in AA analyses}

\author{\speaker{Akihiko Monnai}\\
        KEK Theory Center, Institute of Particle and Nuclear Studies, 
	High Energy Accelerator Research Organization, Tsukuba, Ibaraki 305-0801, Japan\\
        E-mail: \email{akihiko.monnai@kek.jp}}

\abstract{Direct photons are a powerful tool for elucidating the properties of the hot QCD matter in heavy-ion collisions. They are conventionally estimated by assuming prompt photon contributions in proton-proton collisions and thermal and prompt photon contributions in heavy-ion collisions. On the other hand, there could also be other sources of photons such as pre-equilibrium photons. I investigate prompt, pre-equilibrium and thermal photons and their effects on the direct photon $p_T$ spectra at CERN Large Hadron Collider energies.}

\FullConference{International Conference on Hard and Electromagnetic Probes of High-Energy Nuclear Collisions\\
		30 September - 5 October 2018\\
		Aix-Les-Bains, Savoie, France}

\begin{document}

\section{Introduction}

Relativistic nuclear collisions have been a unique tool for investigating hot and dense QCD. One of the important discoveries at BNL Relativistic Heavy Ion Collider (RHIC) is that the system follows hydrodynamic description in the vicinity of the crossover transition between the hadronic and the quark-gluon plasma (QGP) phases. This is supported by the fact that the observed large azimuthal momentum anisotropy, primarily characterized by the elliptic flow $v_2$, is large and relative to the geometrical anisotropy of the system in off-central collisions. The fluidity implies that the QCD matter is thermalized and hadronic distributions are mostly determined at thermal and chemical freeze-outs and subsequent hadronic transport processes. 

On the other hand, photons and dileptons are expected to retain information regarding the spacetime evolution of the system since they do not strongly interact with the medium after their production. Those observables, which are called electromagnetic probes, have been extensively studied to elucidate the detailed properties of the QCD matter created in nuclear collisions. A successful example is the measurements of direct photon particle spectra for experimental determination of the average medium temperatures at RHIC and Large Hadron Collider (LHC), which are shown to be larger than the crossover temperature estimated in lattice QCD, providing an evidence that the produced medium is in the QGP phase \cite{Adare:2008ab,Wilde:2012wc}. Here direct photons are defined as inclusive photons without decay photon contributions.
The quantitative description of the photons, however, is still not completely understood. The theoretical estimations of particle spectra and elliptic flow based on hydrodynamic models are known to systematically underestimate the experimental data. The latter is known as ``photon $v_2$ puzzle" \cite{Adare:2011zr}.

In this study, I estimate non-prompt photon contributions in proton-proton ($pp$) collisions and their effect in heavy-ion ($AA$) analyses for a more complete picture of the direct photon production in nuclear colliders. The conventional model of photon estimation expects that direct photons in $pp$ collisions are prompt photons, which are produced in the hard processes at the collision, and assumes that thermal photons, which are soft radiation from the medium, and prompt photons are the components of direct photons in $AA$ collisions. This picture, however, may not be complete; recent hydrodynamic analyses hint the existence of primordial collective dynamics in small systems \cite{Weller:2017tsr, PHENIX:2018lia} which is under hot debate \cite{Mace:2018vwq}. There is a possibility of thermal photon radiation in most high-multiplicity events of $pp$ collisions \cite{Shen:2016zpp}. Direct photons should be emitted in the pre-equilibrium stage of both $pp$ and $AA$ collisions since the color glass condensate picture implies that glasma production is likely at high energies when the saturation momentum $Q_s$ is sufficiently large. 

\section{The model}

I consider the contributions of thermal photons, pre-equilibrium photons  \cite{Berges:2017eom, Khachatryan:2018ori} and prompt photons for direct photons in nuclear collisions. There can be other sources of photons such as jet photons which shall be discussed elsewhere.

Thermal photons are estimated using a (2+1)-dimensional hydrodynamic model. The equation of state is constructed by matching the one from lattice QCD to the one from the hadron resonance gas model at lower temperatures. The initial condition is based on the Monte-Carlo Glauber model. For demonstrative purposes, I use event-averaged initial conditions at the impact parameter $b = 4.6$~fm for Pb-Pb collisions at $\sqrt{s_{NN}} = 2.76$~TeV. This corresponds to the average impact parameter of the $0$-$20 \%$ centrality events. The nucleon-nucleon inelastic cross section is $\sigma_{pp}^\mathrm{in} = 65$ mb. The thermalization time is chosen as $\tau_\mathrm{th} = 0.4$ fm/$c$.

The thermal photons emission rate is estimated by matching the QGP and the hadronic emission rates as
\begin{eqnarray}
E \frac{dR^\gamma}{d^3p} &=& \frac{1}{2}\bigg(1- \tanh \frac{T-T_c}{\Delta T} \bigg) E \frac{dR^\gamma_\mathrm{had} }{d^3p} + \frac{1}{2}\bigg(1+ \tanh \frac{T-T_c}{\Delta T}\bigg) E \frac{dR^\gamma_\mathrm{QGP}}{d^3p} ,
 \end{eqnarray}
 where $T_c = 0.17$ GeV and $\Delta T = 0.1 T_c$. The QGP rate is based on the perturbative QCD result \cite{Arnold:2001ms} with $N_f = 3$. The QCD coupling is $\alpha_s = 0.2$. The hadronic rate is estimated by taking into account the processes in Refs.~\cite{Turbide:2003si,Heffernan:2014mla,Holt:2015cda}. The thermal photon contributions from the medium before the kinetic freeze-out $T_f = 0.14$ GeV are used.
 
Pre-equilibrium photons are estimated based on the turbulent thermalization approach to glasma. The glasma phase is divided into three stages in the bottom-up scenario \cite{Baier:2000sb} as
\begin{eqnarray}
&\mathrm{(i)}& c_0 Q_s^{-1} \ll \tau \ll c_1 Q_s^{-1} \alpha_s^{-3/2}, \\
&\mathrm{(ii)}& c_1 Q_s^{-1} \alpha_s^{-3/2} \ll \tau \ll c_2 Q_s^{-1} \alpha_s^{-5/2}, \\
&\mathrm{(iii)}& c_2 Q_s^{-1} \alpha_s^{-5/2} \ll \tau \ll c_3 Q_s^{-1} \alpha_s^{-13/5}.
\end{eqnarray}
Here $Q_s$ is the measure of transverse momentum scale implied from the color glass condensate model mentioned earlier. Since the typical time scale of the bottom-up thermalization would be longer than the actual thermalization time implied by the hydrodynamic modeling of nuclear collisions, the auxiliary coefficients $c_0$, $c_1$, $c_2$ and $c_3$ are introduced to scale the processes into the pre-hydrodynamic phase $1/Q_s < \tau < \tau_\mathrm{th}$. The emission rate is based on the Berges-Reygers-Tanji-Venugopalan model \cite{Berges:2017eom,Tanji:2017suk}. In the stages (i) and (ii), it reads
\begin{eqnarray}
E \frac{dR^\gamma}{d^3p} &=& \frac{20}{9\pi^2} \alpha_\mathrm{EM} \alpha_s \log \bigg(1+\frac{2.919}{g^2}\bigg) f_q(p) \int \frac{d^3p'}{(2\pi)^3} \frac{1}{p'} [f_g(p') +  f_q(p')].
 \end{eqnarray}
The quark distribution is parametrized with self-similar scaling as
\begin{eqnarray}
f_q &=& (Q_s \tau)^{-2/3} f_s (p_T,(Q_s \tau)^{1/3}p_z), \\
f_s (p_T,p_z) &=& A p_T^{-1} \exp(-p_z^2/\sigma_z^2),
\end{eqnarray}
where $f_s$ is exponentially cut off for $p_T > Q_s$ and normalized so that the quark number density matches the phenomenological estimation $n_q = c N_f Q_s^3/2\pi^2 Q_s \tau$ where $c = 1.1$. The spatial integral is performed assuming pure longitudinal boost-invariant expansion. In the stage (iii), the emission rate is much closer to the one in equilibrium \cite{Kapusta:1991qp}:
\begin{eqnarray}
E \frac{dR^\gamma}{d^3p} &=& \frac{5}{9} \frac{\alpha_\mathrm{EM} \alpha_s}{2\pi^2} T_\mathrm{eff}^2 \exp (-E/T) \log \bigg(1+\frac{2.919}{g^2}\bigg),
 \end{eqnarray}
where $T_\mathrm{eff} = T_\mathrm{th} \tau/\tau_\mathrm{th}$. Here $T_\mathrm{th}$ is the temperature at the time of thermalization. The emission rate is not the same as the QGP emission rate introduced for the thermal photon estimation because only the Compton scattering and the pair-annihilation processes are considered here. Quantitative discussion should be made carefully because the model is parametrically extended to fit into the pre-hydrodynamic timescales possibly beyond the applicability of the original theory.

Prompt photons are estimated by using the $pp$ direct photon results scaled by the number of collisions $N_\mathrm{coll}$ \cite{Turbide:2003si},
\begin{eqnarray}
E \frac{dR^\gamma}{d^3p} &=& 6745 \frac{\sqrt{s}}{(p_T)^5} \frac{N_\mathrm{coll}}{\sigma_{pp}^\mathrm{in}},
 \end{eqnarray}
where $\sigma_{pp}^\mathrm{in}$ is in the unit of pb for the parametrization. The actual prompt photon contributions may be estimated by subtracting non-prompt photon contributions from total direct photons in $pp$ collisions before the scaling. It should be noted that the low-momentum direct photon spectra of $pp$ collisions is not well-known at LHC energies. One may alternatively use perturbative QCD for the estimation of prompt photons, but should also be careful of its applicability at lower $p_T$. 

\section{Numerical results}

\begin{figure}[tbh]
\includegraphics[width=.5\textwidth]{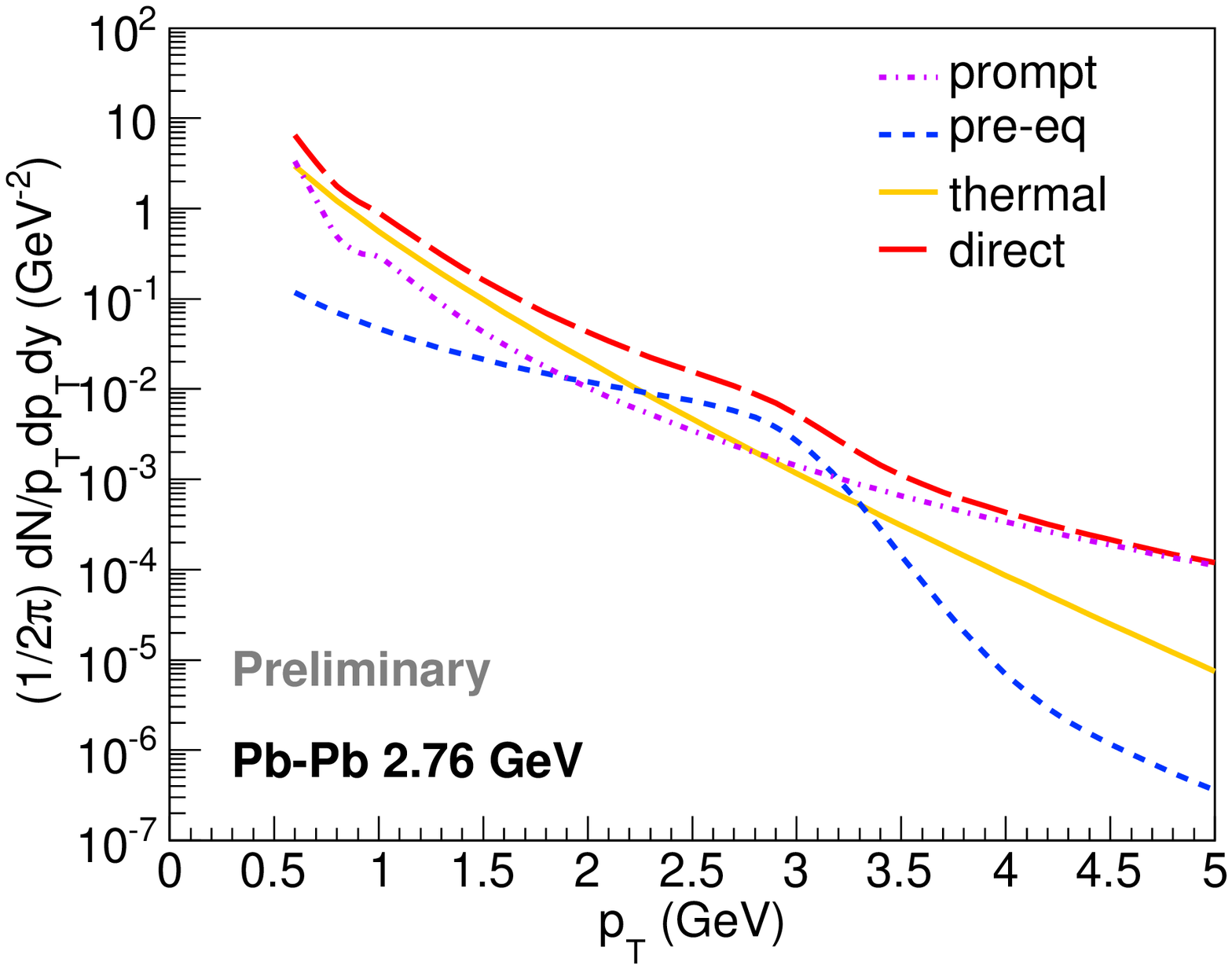}
\includegraphics[width=.5\textwidth]{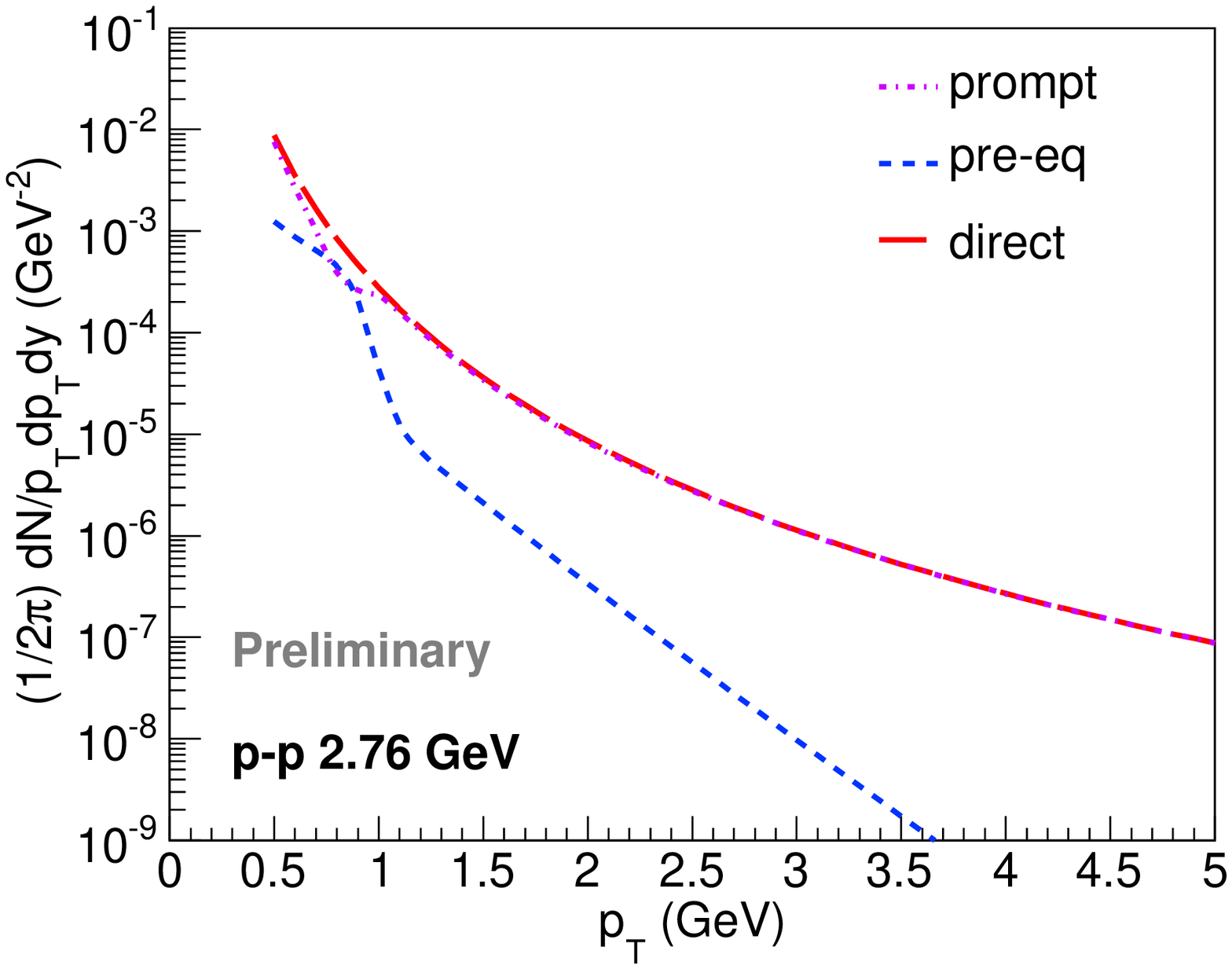}
\caption{Direct photon transverse momentum spectra by components for (left) Pb-Pb collisions and (right) $pp$ collisions at 2.76 TeV.}
\label{fig1}
\end{figure}

The direct photon spectra for the off-central Pb-Pb collisions at $b=4.6$ fm/$c$ at 2.76 TeV are shown in Fig.~\ref{fig1} (left). The pre-equilibrium photons are semi-hard reflecting the saturation momentum scale $Q_s$, which is set to $Q_s = 3$~GeV here. Also the characteristic momentum scale exhibits a time scale ordering from high to low momentum; high $p_T$ regions are dominated by prompt photons, intermediate $p_T$ regions by pre-equilibrium photons and low $p_T$ regions by thermal photons. It should be noted that since pre-equilibrium evolution is not fully understood, the results are still quantitatively dependent on parametrization. 

Figure~\ref{fig1} (right) shows the direct photon spectra for $pp$ collisions. Thermal photon results are not shown here but will be discussed elsewhere since it is still controversial whether such medium can be created in proton collisions. Instead, I consider pre-equilibrium photon contributions and find that it could be comparable to the total photon yields near $Q_s$, which is chosen as $Q_s = 0.9$~GeV. For a given direct photon spectrum, one has to subtract the contribution of non-prompt photons before scaling it with the number of collisions. The spectrum is consequently expected to be reduced near the $Q_s$ of $pp$ collisions and enhanced near that of Pb-Pb collisions.

\section{Summary and outlook}

Direct photon productions in Pb-Pb and $pp$ collisions are studied. Prompt, pre-equilibrium and thermal photons are considered as the components of direct photons. Numerical estimations indicate that pre-equilibrium photons could provide visible semi-hard contribution to $p_T$ spectra. It is implied that $pp$ direct photons may not be pure prompt photons and thus the baseline for AA analyses may well be modified if experimental data of $pp$ collisions are to be used for the scaling. Since pre-equilibrium photons reflects the saturation momentum scale $Q_s$, one may probe and constrain the pre-equilibrium physics experimentally through direct photon analyses. Modification of prompt photons based on color glass condensate has been discussed recently in Ref.~\cite{Benic:2018hvb}. 

Future prospects include the analyses on direct photon elliptic flow $v_2$ to investigate how the pre-equilibrium photons affect the photon puzzle. Full $pp$ thermal photon analyses can be used to test if the QGP is produced in $pp$ collisions experimentally. 
Implementation of the chemically equilibrating QGP in hydrodynamic models may also be important for understanding pre-equilibrium and thermal photons interactively \cite{Gelis:2004ep,Monnai:2014kqa}.

\end{document}